\documentclass{article}
\usepackage{ijcai21}

\usepackage{times}

\usepackage{soul}
\usepackage{url}
\usepackage[small]{caption}
\usepackage{graphicx}
\usepackage{amsmath}
\usepackage{booktabs}
\usepackage{booktabs} 

\usepackage[english]{babel}
\usepackage{moresize}
\usepackage{amsmath}

\usepackage{algorithmic} 
\usepackage{balance}
\usepackage{comment}
\usepackage{paralist}
\usepackage{multirow}
\usepackage{arydshln}
\usepackage{filecontents}

\usepackage[english]{babel}
\usepackage[latin1]{inputenc}
\usepackage{url}
\usepackage{subfigure}
\usepackage{amsmath}
\usepackage{enumitem}
\usepackage[linesnumbered,algoruled,boxed,lined]{algorithm2e}
\usepackage{adjustbox}
\usepackage{amssymb}
\usepackage{times}
\usepackage{hyperref}

\usepackage{pgfplots}
\usetikzlibrary{pgfplots.dateplot}

\usepackage{filecontents}
\definecolor{tblue}{RGB}{31,119,180}
\definecolor{torange}{RGB}{255,127,14}
\definecolor{tgreen}{RGB}{44,160,44}
\definecolor{tred}{RGB}{214,39,40}
\definecolor{tpurple}{RGB}{148,103,189}

\usepackage{filecontents}

\newcommand{\hide}[1]{} 

\newcommand{\etal}{\textit{et al}.}

\newcommand{\ie}{\textit{i}.\textit{e}.}
\newcommand{\eg}{\textit{e}.\textit{g}.}

\title{Recent Advances in Heterogeneous Relation Learning for Recommendation}

\author{Chao Huang\\
\affiliations
University of Hong Kong, Hong Kong\\
chaohuang75@gmail.com
}

\begin{document}

\maketitle

\begin{abstract}
Recommender systems have played a critical role in many web applications to meet user's personalized interests and alleviate the information overload. In this survey, we review the development of recommendation frameworks with the focus on heterogeneous relational learning, which consists of different types of dependencies among users and items. The objective of this task is to map heterogeneous relational data into latent representation space, such that the structural and relational properties from both user and item domain can be well preserved. To address this problem, recent research developments can fall into three major lines: social recommendation, knowledge graph-enhanced recommender system, and multi-behavior recommendation. We discuss the learning approaches in each category, such as matrix factorization, attention mechanism and graph neural networks, for effectively distilling heterogeneous contextual information. Finally, we present exploratory outlook to highlight several promising directions and opportunities in heterogeneous relational learning frameworks for recommendation.
\end{abstract}

\section{Introduction}
\label{sec:intro}

In the era of information explosion, recommender systems are designed to help allievate the information explosion and meet users personalized interests~\cite{he2017neural}. For example, recommending products to users on e-commerce platforms (\eg, Amazon, Ebay, Taobao)~\cite{2019online}, or suggesting movies for users on online video sites (\eg, Netflix, Youtube)~\cite{covington2016deep}. To accurately capture user preference, one of most popular recommendation architectures is collaborative filtering (CF), which leverages the historical co-interact patterns of users on items and make recommendations~\cite{mnih2007probabilistic}.

In our computational online services, practical recommender systems are often associated with heterogeneous structural relations, \eg, users' social connections~\cite{wu2019neural,xu2020global}, knowledge graph of items~\cite{cao2019unifying,ma2019jointly}, multiplex user-item interactions~\cite{xia2020multiplex,jin2020multi}. Such heterogeneous relational data provide opportunities for researchers and practitioners to not only understand and investigate user preferences better, but also present challenges to learn insightful information from them. For example, the incorporation of cross-user social relations could help characterize user preference, based on the social influences among users~\cite{coupledsocialaaai,fan2019graph}. In addition, knowledge graph could provide connectivity information to capture the semantic relatedness between items~\cite{zhu2020knowledge,wang2019knowledgekdd}. Hence, distilling and aggregating knowledge from such heterogeneous structural data would benefit recommendation designs with better personalization. 


In the recommendation with heterogeneous relational context, the key challenge lies in the effective modeling of complex dependent structures from a diverse set of heterogeneous relations. In addition to the difficulty of discovering insightful knowledge from heterogeneous data sources, it is a significant challenge to effectively aggregate the heterogeneous knowledge in a unified way, to facilitate various user modeling applications. Existing heterogeneous relational learning frameworks should be able to address two important questions which corresponds to the aforementioned challenges: (1) How to effectively perform the relation representation learning from the heterogeneous structural data? (2) How to perform the information fusion based on the extracted knowledge via automatic machine learning frameworks, and endow the user preference modeling paradigms with heterogeneous context incorporation?

In this survey, we focus on the recent advances and new trends in heterogeneous relation learning for recommendation. A rich body of learning methods have been applied to solve this challenge in recommender systems, based on various techniques, including attention mechanism~\cite{wu2019dual,chen2019social}, graph neural network~\cite{multibehavioraaai,wang2019knowledgekdd}, knowledge base embedding~\cite{cao2019unifying}, and multi-task learning~\cite{chen2020efficient,gao2019learning}. In the following, we break down this investigation into three parts: i) We identify key challenges of capturing heterogeneous relations for modeling user preferences; ii) We categorize the existing solutions into three groups, \ie, social recommendation, knowledge graph-enhanced recommender systems, and multi-behavior recommendation; iii) We point out promising research directions in heterogeneous relation learning for recommendation. To our best knowledge, this paper is the first survey that pays special attention to recent advances in user modeling and personalization with heterogeneous relational context. 

\section{Collaborative Filtering Based Recommendation}
\label{sec:model}

\begin{figure*}[t]
    \centering
    \includegraphics[width=0.98\textwidth]{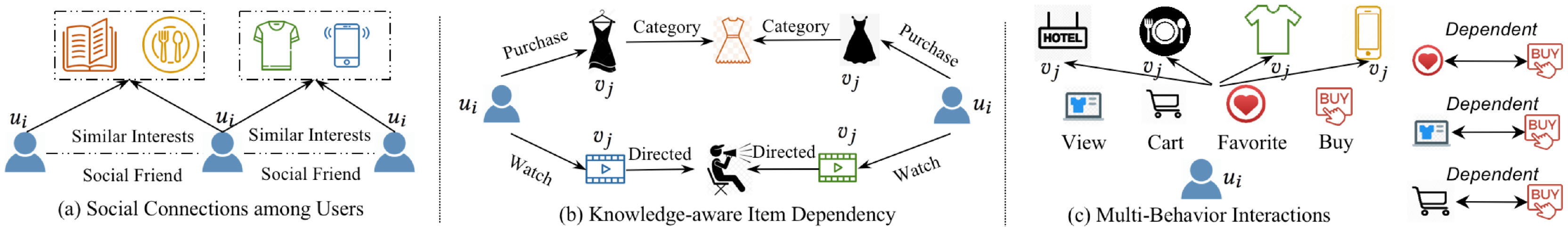}
    \vspace{-0.1in}
    \caption{Illustration of relation heterogeneity learning for recommender systems.}
    \label{fig:examples}
\end{figure*}

Collaborative filtering has emerged as the most popular paradigm to build recommender systems. At its core is to presume that behaviorally similar users likely to share similar interests over items, based on users' historical interaction behaviors (\eg, clicks, purchase)~\cite{he2017neural,hofmann1999latent}. The most common collaborative filtering framework lies in representing users and items with low-dimensional latent features (\ie, embeddings), and then making predictions based on the learned embeddings vectors~\cite{han2018aspect,liang2018variational}.

\subsection{Matrix Factorization}
There exist various types of collaborative filtering techniques which design different model structures for modeling user-item interactions. Specifically, matrix factorization is an early solution to decompose sparse matrix to low-rank latent embeddings, by retaining the dependencies of interactions among users~\cite{mnih2007probabilistic}. For example, Singular Value Decomposition(SVD)~\cite{billsus1998learning} infers the missing elements in user-item rating matrix by factorizing it into a product of two low-rank matrices corresponding to user and item dimensions. Later on, several probabilistic variations are proposed to extend matrix factorization with the integration of the sophisticated Bayesian inference, such as Probabilistic Matrix Factorization (PMF)~\cite{mnih2007probabilistic}, Bayesian Probabilistic Matrix Factorization (BPMF), and Maximum Margin Matrix Factorization (MMMF)~\cite{srebro2005maximum,rennie2005fast}.

In a typical recommendation scenario, we have a set of users $u_i \in \mathcal{U}$ and items $v_j \in \mathcal{V}$ with the size of $I$ and $J$, respectively. The user-item interaction matrix $\textbf{Y}$ is generated based on users' implicit feedback on items, in which $y_{i,j}=1$ if user $u_i$ has interacts with item $v_j$ (\eg, click, watch, or purchase) and $y_{i,j}=0$ otherwise. Matrix factorization models aim to project user and item into low-dimensional latent feature vectors, which can be formally represented as:
\begin{align}
\hat{y}_{i,j} = f(i,j | \Theta)
\end{align}
\noindent where $\hat{y}_{i,j}$ is the predicted probability of interaction $y_{i,j}$ between user $u_i$ and item $v_j$. The learned model parameters are denoted by $\Theta$. $f(\cdot)$ is the encoding function which maps input user interactions $\textbf{Y}$ to the estimated probability $\hat{y}_{i,j}$.

\subsection{Neural Collaborative Filtering Architecture}
With the advent of deep learning in various domains (\eg, computer vision, nature language processing), neural network techniques have been utilized to enhance the collaborative filtering paradigm with non-linear feature interactions. To be specific, NCF~\cite{he2017neural} and DMF~\cite{xue2017deep} leverage multi-layer perceptron to endow the neural collaborative filtering for user-item interaction modeling with non-linearities. In CDAE~\cite{wu2016collaborative} and AutoRec~\cite{sedhain2015autorec}, the autoencoder architecture has been used as the mapping function to learn latent representations with the reconstruction of the user-item interaction input. In addition, inspired from restricted Boltzmann machines~\cite{salakhutdinov2007restricted,larochelle2008classification}, neural autoregressive models are adopted to distribution estimation for collaborative filtering~\cite{zheng2016neural,du2018collaborative}. These neural models perform autoregression for user and item dimensions, so as to capture the inter-correlations between users and items in an explicit way.

Motivated by the development of graph neural network and its strength in relational learning over graph-structured data, many efforts propose to guide the embedding learning over the generated user-item interaction graph~\cite{klicpera2018predict}. Recently emerged graph neural networks (GNNs) design various message passing schemes to encourage neighboring nodes to have similar representations~\cite{kipf2016semi,velivckovic2017graph}. To adapt the GNNs to the collaborative filtering scenario, some studies employ the graph convolutions to aggregate embeddings of neighboring users or items, to refine the representation of the target node, like NGCF~\cite{wang2019neural}, PinSage~\cite{ying2018graph}, DGCF~\cite{wang2020disentangled}, GC-MC~\cite{berg2017graph}.\\

\noindent \textbf{Heterogeneity of Relations}. However, these above collaborative filtering models have thus far focused on representation learning for homogeneous relations for user modeling. As a matter of fact, practical recommendation scenarios involve a diversity of relations from both user and item dimensions~\cite{shi2018heterogeneous} (as shown in Figure~\ref{fig:examples}). Given different recommendation scenarios, methods can be categorized as different dimensions of heterogeneous relation learning:

\begin{itemize}[leftmargin=*]
    \item \textbf{Social Connections among Users}. The increasing popularity of online social platforms enable the connections between users with similar interests, by integrating item recommendations and social network services~\cite{wu2019dual}, \eg, preference of users on items may be influenced by their friends. Hence, one line of heterogeneous relational learning for recommender system is \emph{Social Recommendation} which leverages social information to alleviate the data sparsity and improve the recommendation quality, with the joint modeling of user-user and user-item relations~\cite{jiang2012social,yang2016social}.
    
    \item \textbf{Knowledge-aware Dependencies among Items}. In real-life applications, there exist explicit relations among different items with the consideration of external knowledge (\eg, movie's director, product category, or venue's location)~\cite{xin2019relational}. Such item-wise dependencies can be used to construct the knowledge graph to present the heterogeneous structured information and reflect the semantic relatedness between items. The rich semantics and topology of knowledge graphs could be incorporated into the user embedding process as important side information, by exploiting the knowledge-aware dependency across items~\cite{ma2019jointly}.
    
    \item \textbf{Multi-Behavior User-Item Interactions}. In practice, users exhibit different interaction behaviors on items depending on the intention context~\cite{chen2020efficient}. Let us consider the e-commerce platform as an example: users can click products when they are interested in, or tag these items as favourites if they like them, or make final purchases given the products fulfill their needs~\cite{tanjim2020attentive}. To provide better purchase forecasting for a user, it is beneficial to investigate not only which products he/she has purchased before, but also what this user has browsed and tagged as his/her favorite items~\cite{multibehavior2021}. In such case, user-item interactions are heterogeneous in nature, involving different types of user behaviors (\eg, click, tag-as-favorite, purchase).
\end{itemize}

To provide a better understanding, Figure 1 presents the illustrated examples corresponding to the above recommendation scenarios with heterogeneous relational context.

\section{Heterogeneous Contextual Relation Learning for User Representation}
\label{sec:relate}

In this section, we review the recent advances for user modeling and personalization with the modeling heterogeneous relational context from three perspectives (stated in Section~\ref{sec:model}).

\subsection{Social Recommendation}
Social recommendation aims to probe social relations among users into user-item interaction modeling paradigm, based on the observed phenomena that users' preference is likely to be affected by their social connected friends~\cite{tang2013social}. In social-aware recommender systems, there exist two types of relations: user-user social connections (\eg, friendship, or family) and user-item interactions (\eg, click, or purchase).

\noindent \textbf{Challenge}. The key challenge of social recommendation lies in the effectively transforming the user-user and user-item relational structure into latent factor spaces, such that the semantics of heterogeneous relations can be well preserved.\\

\noindent \textbf{Social-aware Factorization Methods}. In light of the above challenge, the exploration of social recommendation can date back to the development of social relation-enhanced matrix factorization model. In particular, SoReg~\cite{ma2011recommender} considers social signals as the regularization term to constrain the optimized objective in the matrix factorization architecture. Jamali~\etal~\cite{jamali2010matrix} proposes to augment the matrix factorization with the trust influence modeling, so as to capture the social-aware user preference. Ma~\etal~\cite{ma2008sorec} develops a factor analysis method--SoRec which is built on the probabilistic matrix factorization, to jointly investigate users' social connections and interactions. Furthermore, a social collaborative filtering model named TrustMF~\cite{yang2016social} is developed to use the trust relations among users in addressing the data sparsity issue of traditional recommender systems. Nevertheless, those social-aware factorization methods model the two-dimensional interaction of user and item hidden factors, with the assumption of the linear relations between each dimensional latent space. Hence, their model performance can be hindered by the simply linear combination of latent factors~\cite{he2017neural}. \\

\noindent \textbf{Attentive Social Recommender Systems}. Attention mechanism has shown its effectiveness in a variety of tasks, \eg, machine translation~\cite{indurthi2019look} and human activity recognition~\cite{ma2019attnsense}. The key idea of attention mechanism is to identify the influential parts with the attentive weights from the input data. Along with this line, several studies seek to capture the relationships between users based on neural attentive architectures. For instance, Chen~\etal~\cite{chen2019social} proposes a two-stage attentional model (SAMN) to differentiate users' social relations. In SAMN framework, an aspect-level attention module is firstly introduced to generate the user-friend specific relation representation, and then another attention layer is proposed to capture the friend-level differences. Additionally, with the introduction of attention mechanism, an adaptive transfer learning scheme is designed to model the interplay between user and item domain for social recommendation~\cite{chen2019efficient}. To capture social contextual signals, Wu~\etal~\cite{wu2019hierarchical} proposes a hierarchical attention approach to model factors (\ie, past behaviors, social relations, user admiration) which impacts user preferences. \\

\begin{figure}[h]
    \centering
    \includegraphics[width=0.49\textwidth]{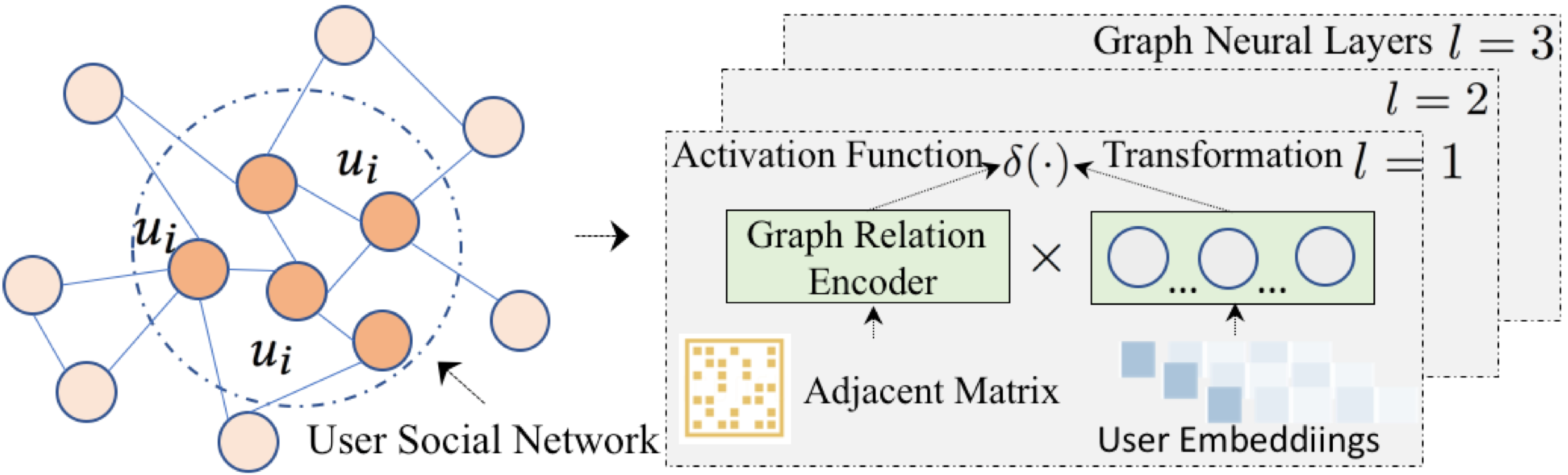}
    \caption{Graph neural network for social recommendation with relation aggregation under message passing paradigm.}
    \label{fig:examples_social_recommendation}
\end{figure}

\noindent \textbf{Graph Neural Network for Social Recommendation}. Advances in graph neural networks (GNNs) have shed light on the development of social recommendation solutions~\cite{coupledsocialaaai}, due to the strong capability of learning on graph structure data. Relations in social recommendation can be naturally represented as two explicit graph-based heterogeneous structures: i) user-item interactions; ii) user-user social connections. Along with this line, many efforts have developed graph neural networks with information propagation paradigm, to inject social signals into the user-item interaction modeling. We show the overall model flow of graph neural network-based social recommender systems in Figure~\ref{fig:examples_social_recommendation}. In general, current methods can be summarized into the following graph neural paradigms. \\

\noindent (i) \textbf{Graph Convolutional Network-based Models}. Inspired by the prevalence of spectral graph learning framework--Graph Convolutional Network (GCN), DiffNet~\cite{wu2019neural} simulates the influence process between users, by recursively propagating information over the social network for user and item representation refinement. The designed influence diffusion neural layers can be incorporated into the collaborative filtering architecture, to capture the heterogeneous relational context for social-aware recommendation. In RecoGCN~\cite{xu2019relation}, the GCN is adopted to aggregate heterogeneous features over the constructed heterogeneous graph with the semantic-aware meta-paths from user-user and user-item edges. To address the sparsity problem, a GCN-based autoencoder is introduced to augment the data with the adversarial learning strategy~\cite{yu2020enhance}. \\

\noindent (ii) \textbf{Graph Attention Network-based Models}. To discriminate the importance of neighbouring feature aggregation, Graph Attention Network (GAT)~\cite{velivckovic2017graph} is proposed as a non-spectral learning method that leverages the node spatial information for embedding propagation. To alleviate the limitation of the mean-based information aggregator, GraphRec~\cite{fan2019graph} integrates attention mechanism into the graph neural network to model heterogeneous user latent factors from social space. To endow the social recommender system to capture multifaceted social relations, Wu~\etal~\cite{wu2019dual} devises a dual-stage graph attention to dynamic weigh social effects from user and item domain context. To inject temporal context into social recommendation, GAT is adopted to cooperate with a recurrent neural network to capture the time-aware social influence and user interests~\cite{song2019session}. \\

\noindent \textbf{General GNN Social Recommendation Framework}. Suppose $\textbf{H}^l[u_i]$ is the representation of user $u_i$ at the $(l)$-th GNN layer, the message passing paradigm from the $(l)$-th layer to the $(l+1)$-th layer can be formally represented as:
\begin{align}
\textbf{H}^{(l+1)}[u_i] = \text{Aggre} \Big ( \text{Extract} (\textbf{H}^l[u_i], \textbf{H}^l[v_j]), \textbf{H}^l[u_{i'}] ) \Big )
\end{align}
\noindent where $N_{u_i}^s$ and $N_{u_i}^v$ denote the set of user $u_i$'s social connected users and items, respectively. $u_{i'} \in N_{u_i}^s$; $v_{j} \in N_{u_i}^v$. The key GNN operators: Extract($\cdot$) represents the heterogeneous neighbor information extractor from user-user and user-item dependent context. Aggre($\cdot$) is the aggregation function to fuse heterogeneous signals, including the mean and attention-based operator corresponding to the GCN and GAT message passing methods, respectively. \\

\noindent \textbf{Summary}. To adapt to the heterogeneous relational learning scenarios, all existing social recommendation models employ different structural information aggregation schemes, \eg, attention mechanism, graph convolution and graph attention network. The nature of graph neural networks is able to jointly maintain user-user and user-item relational structures, for generating contextualized user and item embeddings.




\subsection{Knowledge Graph-enhanced Recommendation}
Knowledge Graphs (KGs) brings up a large amount of commercial benefits for various applications, ranging from question answering~\cite{yang2019knowledge} and information extraction~\cite{yang2019transms}. KGs provide world's factual information with heterogeneous relational structures (entity-relation-entity triplets), in which nodes and edges represent the entities and the corresponding entity-wise relations, respectively. In recent years, incorporating knowledge graph information into the recommendation has attracted much attention, to help improve the performance of recommender systems~\cite{zhu2020knowledge}. Such knowledge side information presents heterogeneous relations corresponding to different item attributes, which i) captures the rich semantic relatedness among items; ii) improves the explainability of recommender systems with knowledge graph connections.

\noindent \textbf{Challenge}. The main challenge in incorporating knowledge graph into recommendation lies in how to capture the semantic relatedness between items, and transform such underlying dependencies into the user/item representation paradigm with the heterogeneous relational context. \\

To address the aforementioned challenge, existing knowledge graph-enhanced recommendation techniques can be grouped into three main types: path-based approaches, regularization-based methods, and hybrid learning methods. \\\vspace{-0.12in}

\begin{figure}[h]
    \centering
    \includegraphics[width=0.40\textwidth]{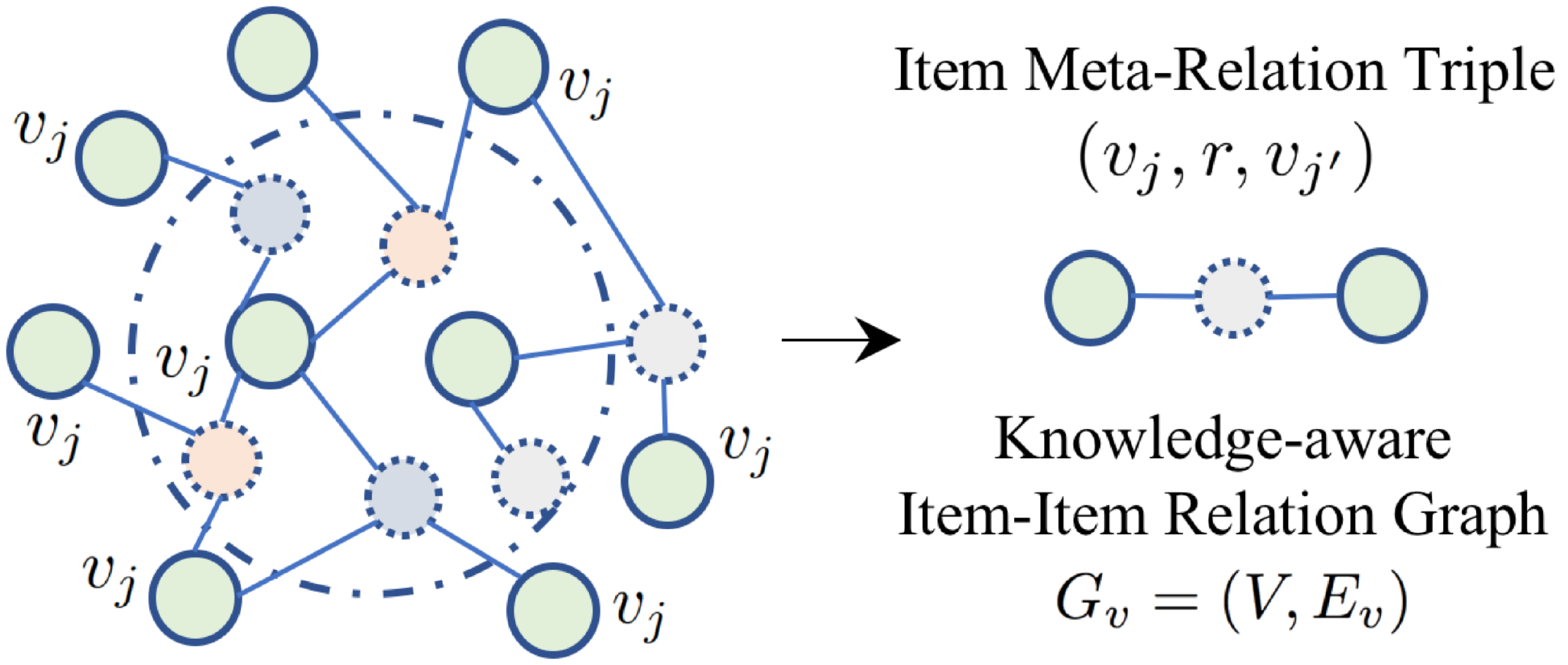}
    \caption{Knowledge graph-guided item meta-relations.}
    \label{fig:examples_knowledge_graph}
\end{figure}

\noindent \textbf{Path-based Approaches}. The path-based approaches attempt to generate entity paths between graph nodes based on their high-order connections (as illustrated in Figure~\ref{fig:examples_knowledge_graph}). Early personalized entity recommendation method~\cite{yu2014personalized} considers the meta-path heterogeneous graph as external knowledge to alleviate the data scarcity issue. Motivated by the deep learning methods over heterogeneous information network, a co-attention mechanism is proposed to inject the meta-path based context into the recommendation architecture, so as to characterize the three-way knowledge-based interaction with the triple form of (user, meta-path, item)~\cite{hu2018leveraging}. To effectively identify useful meta-graph based
features (\eg, user$\rightarrow$review$\rightarrow$business), Zhao~\etal~\cite{zhao2017meta} develops a group lasso regularized factorization machine, to characterize the knowledge-aware semantics. However, the success of most path-based methods largely relies on the effective design of meth paths for modeling heterogeneous relations among users and items, which requires domain-specific knowledge. Hence, such process may hardly be adapted to the complex knowledge graph structures with diverse categories of entities and relations.\\

\noindent \textbf{Regularization-based Methods}. Inspired by the strength of knowledge graph embedding techniques (\eg, TransE~\cite{bordes2013translating}), several regularization-based methods propose to regularize the user representation learning in recommendation with the designed additional knowledge graph loss. Following this line, regularization-based methods integrate the tasks of knowledge graph learning and recommendation under a multi-task learning paradigm. For example, KTUP~\cite{cao2019unifying} simultaneously models item recommendation and knowledge graph completion, to learn user preference with the consideration of knowledge-aware relations. To explain the recommendation results, Ai~\etal~\cite{ai2018learning} generates the interpretations based on embedded knowledge base, with the preservation of heterogeneous dependent structures. Due to the lack of modeling high-order connectivity between users and items, these methods cannot capture the long-range dependencies for user-item interactions and knowledge graphs. \\

\noindent \textbf{Hybrid Learning Methods}. Another line of KG-enhanced recommendation models learn user/item representations with the joint modeling of i) path-based entity dependence; and ii) knowledge graph structural semantics. For instance, KARN~\cite{zhu2020knowledge} proposes an attentive recurrent network to incorporate knowledge graph into the sequence learning framework of user's click records. A joint optimization framework is developed to extract inductive rules from item-centric knowledge graphs~\cite{ma2019jointly}. In addition, graph neural networks have been utilized for encoding the knowledge-aware relations to transform the knowledge graph into user interest modeling, such as KGNN-LS~\cite{wang2019knowledgekdd} and KGAT~\cite{wang2019kgat}. In particular, KGNN-LS is a knowledge-aware graph neural model with the integration of label smoothness regularization. This work is on the basis of the assumption that users are likely to share similar preferences on items which are connected in the knowledge graph. KGAT enables the high-order knowledge-aware relation learning under the graph attention architecture. \\

\noindent \textbf{Summary}. Overall, the limitation of path-based methods lies in the requirement of domain-specific knowledge for effective meta-path construction, which hinders the flexibility of those KG-enhanced recommendation solutions. The success of regularization-based methods replies on the output of knowledge base embedding. To effectively leverage external knowledge, it is important to design a main representation space which effectively bridges the recommendation and knowledge graph learning. Compared with the path-based approaches and regularization-based methods, the hybrid learning models have a better capability with respect to discovering complex knowledge-aware heterogeneous relations. However, by involving the recursive embedding propagation, GNN-based methods have large computational cost by exploring the high-order connectivity among entities. Hence, how to achieve a nice trade-off between the modeling of long-range collaborative knowledge and model efficiency, remains a significant challenge.

\subsection{Multi-Behavior Recommender System}
In real-world applications, users often interact with items based on various types of behaviors, \eg, click, tag-as-favorite, add-to-cart and purchase in online retailing systems~\cite{xia2020multiplex}. Hence, it is important to capture the collaborative signals from multi-behavioral patterns, to characterize diverse user-item relationships. In the behavior-aware recommendation settings, the behavior type we aim to forecast is regarded as \emph{target
behavior}, such as purchase in online retailing platforms or positive rating feedback in online review sites. Other types of user behavior are regarded as \emph{auxiliary behavior}. For example, in e-commerce platforms, both add-to-cart and tag-as-favorite behavior could serve as indicators to model the purchase patterns of users.

\noindent \textbf{Challenge}. In the multi-behavior recommendation scenario, different types of behavior may reflect the interactive pattern between users and items from different perspectives. It is important to capture the unique context and semantics of heterogeneous user-item interactions when modeling the multi-behavioral relationships~\cite{multibehavioraaai}.Because diverse types of user-item interactions can be dependent in a complex way, it is important to design an effective learning model to capture the underlying multi-behavior dependencies.


Existing multi-behavior recommendation methods can be categorized into two key dimensions: multi-task behavior learning frameworks and graph-based recommender systems. \\


\noindent \textbf{Multi-Task Behavior Learning Frameworks}. The first type of studies tackle the multi-behavior recommendation under a multi-task learning architecture. Particularly, the modeling of multi-typed behavioral interactions is conducted with multiple tasks based on the shared embedding space. For example, To account for multi-typed user behaviors, Gao~\etal~\cite{gao2019learning} integrates the neural collaborative filtering, to learn separate interaction encoding function for individual type of behaviors under a multi-task learning paradigm. Furthermore, in EHCF~\cite{chen2020efficient}, a transfer-based model is developed to correlate the forecasting of each type of user-item interaction. With the designed transfer learning approach, EHCF can not only capture heterogeneous relational structure between user and item, but also alleviate the data sparsity issue for inactive users. \\

\begin{figure}[h]
    \centering
    \includegraphics[width=0.43\textwidth]{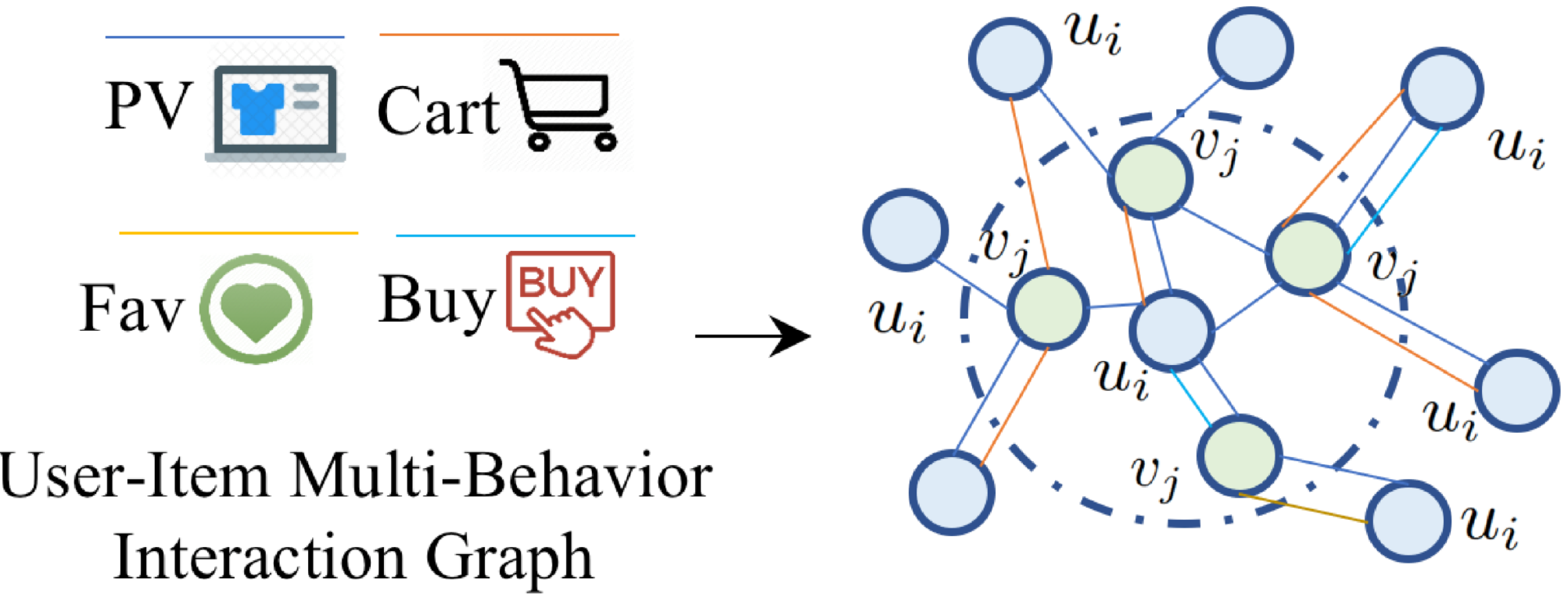}
    \caption{Multi-behavior user-item interaction graph.}
    \label{fig:examples_multibehavior_recommendation}
\end{figure}

\noindent \textbf{Graph-based Multi-Behavior Recommender Systems}. The second type of multi-behavior recommender systems are built upon the graph neural architectures. These methods firstly construct the heterogeneous user-item interaction graph by differentiating behaviors (as shown in Figure~\ref{fig:examples_multibehavior_recommendation}). Edges in the generated graph represent different types of user behaviors (\eg, click, tag-as-favorite, purchase). Then, they leverage different graph relation encoders (\eg, graph convolutional network and graph attention mechanism), to aggregate multi-relational contextual signals during the message passing procedure. For example, MBGCN~\cite{jin2020multi} is a behavior-aware information propagation framework based on graph convolutional operations, which captures the semantics of multi-typed behaviors. Through extensive experiments, authors demonstrate the performance superiority of multi-behavior models as compared to the recommender systems with singular type of behaviors.

To effectively and explicitly discriminate the influences of different types of behaviors, KHGT~\cite{multibehavioraaai} builds a graph-structured transformer network to determine which types of user-item interactions contribute more in assisting the prediction of target behavior (\eg, purchase).
In those graph-based multi-behavior recommendation methods, the behavior-aware embeddings are propagated over the multi-behavior interaction graph to refine user and item representations with the high-order relational structures. \\

\noindent \textbf{Summary}. Different from the heterogeneous relations extracted from user or item domain as side information (\eg, social connections, item knowledge-aware relations), edges between node pairs in the multi-behavior recommendation present the relation multiplicity. Such multiplex topological structures between user and item are investigated in the recently emerged multi-behavior recommender systems.

\section{Future Challenges and Opportunities}
\label{sec:future_work}



This section presents a few potential opportunities for recommendation with heterogeneous relational context.


\begin{itemize}[leftmargin=*]
\item \textbf{Learning to Recommend with Multi-Modal Representations}. The prevalence of web services allow the recommender systems to access online multi-modal content, such as item images/videos or textural description, user profile and reviews~\cite{wei2020graph}. Along with this line, learning informative multi-modal representations of users and items with the incorporation of multi-modal content data into heterogeneous relational learning paradigms, become an important theme of the recommendation frameworks.

\item \textbf{Learning with Data Scarcity}. The effectiveness of current methods relies on sufficient relational data, and may not be able to learn quality embeddings when user behavior or heterogeneous relation data are scarce. They often require a sufficient amount of user-item interactions, or knowledge graph connections, so that the encoded representations is able to well capture the dependent structures between entities. However, many real-life systems involve a limited amount of relational data. Therefore, to build effective heterogeneous relational learning models, it is important to develop effective learning architectures to perform the heterogeneous relation learning with data scarcity.


\item \textbf{Temporal Context Injection}. In real-life online platforms, the temporal patterns of user-item interactions and the underlying time-aware behavior relationships, serve as a key feature dimension in many recommender systems~\cite{sessionaaai}. Hence, it is necessary to design dedicated learning frameworks for modeling dynamic heterogeneous relational data with complex temporal context. However, recommendation models with temporal heterogeneous relations still face several challenges: i) from the local perspective, connections among users and items are inherently dynamic; ii) from the global perspective, it is important to capture the temporal long-term structures with the relational heterogeneity, which calls for dedicated efforts.

\item \textbf{Interpretable User Modeling}. The model interpretation ability with heterogeneous relational learning is less explored in current recommender systems. Specifically, most of current approaches only focus on forecasting unknown user-item interactions, but ignore the inference of casual factors for predicted results. In practical recommendation scenarios, identifying the underlying factors which influence user interaction behaviors, is beneficial to show result explanations and further improve the recommendation performance. For example, performing accurate reasoning on cross-type behavior dependent structures to infer user preferences towards target items, will endow the multi-behavior recommender systems with the capacity of providing informative explanations for identifying multi-dimensional factors. Therefore, it would be interesting and desirable to leverage advanced techniques (\eg, causal inference~\cite{pearl2019seven}) and develop interpretable paradigms for heterogeneous relation learning-based recommender systems.




\item \textbf{Real-Time Computational Frameworks}. Notwithstanding the promising performance of existing heterogeneous relation learning-based recommendation frameworks, most of them operate on the static observed user behaviors and side features. This makes those methods less suited to the practical applications, where new heterogeneous relational data (\eg, clicks and purchases of users, dynamic user connections) arrive continually over time. Hence, real-time computational frameworks are important because they allow the heterogeneous relational recommender systems to infer user preference from streaming data.


\end{itemize}





\section{Conclusion}
\label{sec:conclude}

This work introduces the problem of heterogeneous relational learning for recommendation and discuss its corresponding challenges. We present the recent advances in addressing these challenges. This survey not only provides a comprehensive study of user modeling techniques with heterogeneous context for researchers and practitioners in related fields, but also shed light on designing solutions to harness the power of heterogeneous data, and distill useful knowledge for various real-life user modeling and personalization applications.

\clearpage

\bibliographystyle{named}
\bibliography{ijcai21}

\end{document}